\documentstyle[revtex]{aps}
\begin{document}
\draft

\hspace{11cm} LBL-35097
\vspace{1cm}

\begin{title} COLOR, SPIN AND FLAVOR DIFFUSION IN QUARK-GLUON PLASMAS
\footnote{This work was supported in part by DOE grant No. DE-AC03-76SF00098,
and the Danish Natural Science Research Council. }
\end{title}
\vspace{1cm}

\author{H. HEISELBERG}
\vspace{1cm}

\begin{instit}
        Nuclear Science Division, MS 70A-3307\\
        Lawrence Berkeley Laboratory, \\
        University of California, Berkeley, California 94720
\end{instit}

\vspace{3cm}
\begin{center} January, 1993 \end{center}

\newpage
\begin{title}COLOR, SPIN AND FLAVOR DIFFUSION IN QUARK-GLUON PLASMAS\end{title}
\author{H. Heiselberg}
\begin{instit}
        Nuclear Science Division, MS 70A-3307\\
        Lawrence Berkeley Laboratory, \\
        University of California, Berkeley, California 94720
\end{instit}
\receipt{january 1994}
\begin{abstract}

  In weakly interacting quark-gluon plasmas diffusion of color is found to be
  much slower than the diffusion of spin and flavor because color is easily
  exchanged by the gluons in the very singular forward scattering processes.
  If the infrared divergence is cut off by a magnetic mass, $m_{mag}\sim
  \alpha_sT$, the color diffusion is $D_{color}\sim
  (\alpha_s\ln(1/\alpha_s)T)^{-1}$, a factor $\alpha_s$ smaller than spin and
  flavor diffusion. A similar effect is expected in electroweak plasmas above
  $M_W$ due to $W^\pm$ exchanges.  The color conductivity in quark-gluon
  plasmas and the electrical conductivity in electroweak plasmas are
  correspondingly small in relativistic heavy ion collisions and the very
  early universe.

\end{abstract}

\setlength{\oddsidemargin}{1.5cm}
\setlength{\evensidemargin}{\oddsidemargin}
\columnsep -.8cm
\twocolumn
\narrowtext


Transport of color degrees of freedom in a quark-gluon plasma has recently been
found to be infrared sensitive \cite{GS} and thus differ from other transport
processes as viscous and thermal flow, stopping and electrical conduction
\cite{BP1,BP2} as well as energy degradation \cite{GT}.  Color does not flow
easily due to the transfer of color (color-flip) in the exchange of a colored
gluon in the very singular forward collisions.  As will be discussed here, this
suppression of the color flow also applies to the color diffusion and color
conductivity, which are infrared sensitive like the quark and gluon
quasiparticle relaxation rates \cite{damp,HP}.  These results do not only apply
to QCD plasmas consisting of quarks, antiquarks and gluons, as existing in the
Early Universe and searched for in relativistic heavy ion collisions at
Brookhaven (AGS and RHIC) and CERN (SPS and LHC). As will be shown below it is
a general mechanism for most non-abelian gauge theories.  For example, in an
electroweak plasma consisting of leptons, photons and electroweak bosons
at temperatures above $M_W\simeq 80$ GeV, the exchange of $W^\pm$ provides a
charge exchange analogous to the color exchange which results in
correspondingly poor electrical diffusion and conduction.

Earlier studies of
transport processes in relativistic quark-gluon and electron-photon plasmas
found that the effect of Landau damping effectively led to screening of
transverse interactions and gave the characteristic relaxation rates in
transport processes. Transport coefficients for weakly interacting
electron-photon and quark-gluon plasmas for both thermal plasmas
\cite{BP1,BP2,GT} as well as degenerate ones \cite{deg} were calculated to
leading logarithmic order. Generally the transport relaxation rates have the
following dependence on interaction strength
\begin{eqnarray}
  1/\tau_{tr} \sim \alpha_s^2\ln(1/\alpha_s) T \, . \label{tr}
\end{eqnarray}
However, the quark and gluon quasiparticle damping rates, $1/\tau_p$, were not
sufficiently screened by Landau damping for non-vanishing quasiparticle
momentum, ${\bf p}$, and depends on an infrared cut-off,
$m_{mag}\simeq\alpha_sT$, so that \cite{damp,HP}
\begin{eqnarray}
   1/\tau_p^{(g)} = 3\alpha_s\ln(1/\alpha_s) T \, , \label{gg}
\end{eqnarray}
to leading logarithmic order.  Since the quasiparticle decay rates are not
measurable transport coefficients the infrared sensitivity was not considered a
serious problem.  However, it was recently discovered \cite{GS} that diffusion
of color in some abstract color space suffered from the same infrared
divergence which led to the same color relaxation rates as the quasiparticle
damping rates.

We will describe the two kinds of transport processes by calculating
the flavor, spin, and color diffusion coefficients in a quark-gluon plasma
within the Boltzmann kinetic equation
\begin{eqnarray}
   (\frac{\partial}{\partial t}&+&
   {\bf v}_{{\bf p}_1}\cdot\nabla_{\bf r}  +
   {\bf F}\cdot\nabla_{{\bf p}_1} ) n_1  =\,
 - 2\pi\nu_2\sum_{{\bf p}_2{\bf p}_3{\bf p}_4}  \nonumber\\
   &\times&
  [ n_1n_2(1\pm n_3)(1\pm n_4) - n_3n_4(1\pm n_1)(1\pm n_2)] \nonumber\\
 &\times& |M_{12\to 34}|^2
    \delta_{{\bf p}_1+{\bf p}_2 ,{\bf p}_3+{\bf p}_4}
    \delta (\varepsilon_1 +\varepsilon_2 -\varepsilon_3 -\varepsilon_4 )
     .  \nonumber\\
 && \label{BE}
\end{eqnarray}
Here ${\bf p}_i$ and $\varepsilon_i$ are the quasiparticle momentum
and energy respectively, $n_i({\bf p}_i)$ the
quasiparticle distribution function, and ${\bf F}$ the force on a
quasiparticle. The r.h.s. is the collision integral
for scattering particles from initial states 1 and 2 to
final states 3 and 4, respectively with matrix element squared
$|M_{12\to 34}|^2$ summed over final states and averaged over
initial states.
The $(1\pm n_i)$ factors correspond physically to the Pauli blocking of
final states, in the case of fermions, and to (induced or) stimulated
emission, in the case of bosons. $\nu_2$ is the statistical factor,
$\nu_2=16$ for gluons and $\nu_2=12N_f$ for quarks and antiquarks.
For scattering of quarks of different flavor
\begin{eqnarray}
   |M^{(qq')}_{12\to 34}|^2 = \frac{4}{9}  g^4 \, \frac{u^2+s^2}{t^2}
  \frac{1}{16\varepsilon_1\varepsilon_2\varepsilon_3\varepsilon_4}
   \, ;    \label{M}
\end{eqnarray}
quark-gluon and gluon-gluon interactions are just
9/4 and $(9/4)^2$ times stronger respectively near forward scattering.
In a medium this singularity is screened as given by the Dyson equation
in which a gluon self-energy $\Pi_{L,T}$ is added to the propagator
\begin{eqnarray}
    t^{-1} \to \omega^2-q^2-\Pi_{L,T} \, , \label{Dyson}
\end{eqnarray}
(we refer to \cite{We} for details on separating longitudinal
and transverse parts of the interaction)
where the longitudinal and transverse parts of the
self-energy in QED and QCD are for $\omega,q\ll T$ given by
\begin{eqnarray}
  \Pi_L(\omega,q) &=& q_D^2 \left(1-\frac{x}{2}\ln\frac{x+1}{x-1}\right)
  \,  ,\label{PiL} \\
  \Pi_T(\omega,q)  &=& q_D^2 \left[\frac{1}{2}x^2+\frac{1}{4}x(1-x^2)
      \ln\frac{x+1}{x-1}\right] \, , \label{PiT}
\end{eqnarray}
where $x=\omega/qv_p$ and $v_p=c$ for the relativistic plasmas considered here.
The Debye screening wavenumber in thermal QCD is $q_D^2=g^2(2N+N_f)T^2/6$ where
$N=3$ is the number of colors, $N_f$ is the number of quark flavors, $T$ the
plasma temperature and $\mu_q$ the quark chemical potential.  We refer to
\cite{LH} for a detailed comparison to QED plasmas.  In the static limit,
$\Pi_L(\omega=0,q)=q_D^2$, and the longitudinal interactions are Debye
screened.  For the transverse interactions the self-energy obeys the
transversality condition $q^\mu \Pi_{\mu\nu} = 0$, which insures that the
magnetic interactions are unscreened in the static limit,
$\Pi_T(\omega=0,q)=0$.  It has therefore been suggested that the transverse
interactions are cut off below the ``magnetic mass", $m_{mag}\sim g^2T$, where
infrared divergences appear in the plasma \cite{Linde}.  However, as was shown
in \cite{BP1,BP2}, dynamical screening due to Landau damping effectively screen
the transverse interactions off in most transport problems at a length scale of
order the Debye screening length $\sim 1/gT$ as in Debye screening.
Nevertheless, there are three important length scales in the quark-gluon
plasma. For a hot plasma they are, in increasing size, the interparticle
spacing $\sim 1/T$, the Debye screening length $\sim 1/gT$, and the scale
$1/m_{mag}\sim 1/g^2T$ where QCD effects come into play.  A weakly interacting
QCD plasma and its screening properties is very similar to a QED plasma if one
substitutes the fine structure constant $\alpha_s=g^2/4\pi$ by
$\alpha=e^2/4\pi\simeq 1/137$, the gluons by photons and the quarks by leptons
with the associated statistical factors \cite{LH}.

 Let us first consider a quark-gluon plasma where the particle flavors have
been separated spatially, i.e.,
the flavor chemical potential depends on position, $\mu_i({\bf r})$.
In a steady state scenario the quark flavors will then be flowing with
flow velocity, $u_i$. For simplicity we take the standard ansatz for the
distribution functions (see, e.g., \cite{deg,BPbook})
\begin{eqnarray}
   n_i({\bf p}_i) &=&\left(\exp(\frac{\varepsilon_{\bf p}-\mu_i({\bf r})-
                  {\bf u}_i\cdot{\bf p}}{T})\pm 1\right)^{-1} \nonumber\\
    &\simeq&  n_i^0 -
   \frac{\partial n_p}{\partial\varepsilon_p} {\bf u}_i\cdot{\bf p} \, .
   \label{ni}
\end{eqnarray}
The expansion is valid near equilibrium where $\mu_i$ and
therefore also ${\bf u}_i$ is small. It gives
two terms: the equilibrium distribution function
$n_i^0=(\exp[(\varepsilon_{\bf p}-\mu_i({\bf r})/T])\pm 1)^{-1}$
and the
deviation from that. In general the deviation from equilibrium has
to be found self-consistently by solving the Boltzmann equation. However,
as in the case of the viscosity \cite{BP1}, we expect the ansatz (\ref{ni})
to be good within few percent to leading logarithmic order.

The flavor diffusion coefficient, $D_{flavor}$, defined by:
\begin{eqnarray}
  {\bf j}_i = -D_{flavor} \nabla \rho_i  \, ,
\end{eqnarray}
is given in terms of the flavor current ${\bf j}_i$ and the gradient of the
number density $\rho_i=\sum_p n^0_i({\bf p}_i)=\nu_iT^33\xi(3)/4\pi^2$
of a particular flavor {\it i}. From (\ref{ni}) we find
\begin{eqnarray}
   {\bf j}_i = \sum_{\bf p} n_{{\bf p},i} = {\bf u}_i \rho_i \, .
\end{eqnarray}
The density gradient,
$\nabla\rho_i=\nabla\mu_i\sum_p(\partial n^0_i/\partial\varepsilon_p)$
can be found by solving the Boltzmann equation.
Linearizing in ${\bf u}_i$ we obtain
\begin{eqnarray}
    \frac{\partial n_1}{\partial\varepsilon_1} {\bf v}_1\cdot &\nabla\mu_1 &
     = 2\pi\nu_2\sum_{{\bf p}_2{\bf p}_3{\bf p}_4}    |M_{12\to 34}|^2
    \nonumber\\ &\times&
    n^0_1n^0_2(1-n^0_3)(1\pm n^0_4)   \nonumber\\
    &\times& ({\bf u}_1\cdot{\bf p}_1+{\bf u}_2\cdot{\bf p}_2
    -{\bf u}_3\cdot{\bf p}_3-{\bf u}_4\cdot{\bf p}_4)     \nonumber \\
    &\times& \delta_{{\bf p}_1+{\bf p}_2 ,{\bf p}_3+{\bf p}_4}
    \delta (\varepsilon_1 +\varepsilon_2 -\varepsilon_3 -\varepsilon_4 )
    \, . \label{BE2}
\end{eqnarray}
It is most convenient to choose the plasma center-of-mass system where
one flavor is flowing with velocity ${\bf u}_1$ and
the others with velocity ${\bf u}_2=-{\bf u}_1/(N_f-1)$. The number of
scatterers is then $\nu_2=12(N_f-1)$.
Equivalently, one can conveniently include the first flavor
so that the number of scatterers is $\nu_2=12N_f$ but ${\bf u}_2=0$.
In steady state the gluons will not move in the c.m.s., i.e.  ${\bf u}_2=0$
for quark-gluon scattering.
Since the flavor is unchanged in the collisions ${\bf u}_3={\bf u}_1$ and
${\bf u}_4={\bf u}_2$.

To leading logarithmic order the singular interaction near forward scattering
allows us to expand around ${\bf q}\sim 0$,
where ${\bf q}={\bf p}_1-{\bf p}_3={\bf p}_4-{\bf p}_2$ is the momentum
transfer in the collision.
Multiplying both sides of (\ref{BE2}) by ${\bf p}_1$ and summing the
Boltzmann equation reduces to
\begin{eqnarray}
   n_1\nabla\mu_1
   &=& -{\bf u}_1\frac{\pi}{3}\nu_2\sum_{{\bf q},{\bf p}_1,{\bf p}_2}
      n^0_1n^0_2(1-n^0_3)(1\pm n^0_4)  \nonumber\\
  &\times&  |M_{12\to 34}|^2 q^2
    \delta (\varepsilon_1 +\varepsilon_2 -\varepsilon_3 -\varepsilon_4 )
    \, , \label{BE3}
\end{eqnarray}
where we have used the antisymmetry of the r.h.s. by coordinate change
${\bf p}_1\to {\bf p}_3$ so that ${\bf p}_1\to {\bf q}/2$.
The r.h.s. collision integral of Eq. (\ref{BE3})
is now straightforward to evaluate to leading logarithmic order when the
screening is properly included (see also Refs. \cite{BP1,BP2,deg}). We find
\begin{eqnarray}
    D_{flavor}^{-1} \simeq  \frac{\pi^5}{3^34\xi(3)^2} (1+N_f/6)
        \alpha_s^2\ln(1/\alpha_s) T \, ,\label{Di}
\end{eqnarray}
where $\xi$ is the Rieman zeta-function.
The term ``1" arises from quark-gluon scatterings and the $N_f/6$ from
quark-quark scatterings.
 This result is similar to the viscous, thermal and momentum
relaxation rates because the collision term contains the same factors of
momentum transfer: the singular $q^{-4}$ factor
from the matrix element squared and
the suppressing $q^2$ factor because the quark flavors lose little momentum
in forward scatterings. Including screening, $q^{-4}\to (q^2+\Pi_{L,T})^{-2}$,
where effectively $\Pi_{L,T}\sim q_D^2$,
and integrating over momentum transfer, $d^2q$, gives the leading logarithmic
term $\ln(T^2/q_D^2)\simeq \ln(1/\alpha_s)$.

  Subsequently, let us consider the case where the particle spins have
been polarized spatially by some magnetic field \cite{BPbook}, i.e.,
the spin chemical potential depends on position, $\mu_\sigma({\bf r})$.
With the analogous ansatz to (\ref{ni}) for the distribution functions with
$\mu_\sigma$ instead of $\mu_i$,
we find the spin current ${\bf j}_\sigma={\bf u}_\sigma n_\sigma$.
Linearizing the Boltzmann equation we find
\begin{eqnarray}
   \frac{\partial n_1}{\partial\varepsilon_1}
    \nabla\mu_{1,\sigma}\cdot &{\bf v}_1&
   = 2\pi\nu_2\sum_{{\bf p}_2{\bf p}_3{\bf p}_4}
    n^0_1n^0_2(1-n^0_3)(1\pm n^0_4)  \nonumber\\ &\times&
     \delta_{{\bf p}_1+{\bf p}_2 ,{\bf p}_3+{\bf p}_4}
    \delta (\varepsilon_1 +\varepsilon_2 -\varepsilon_3 -\varepsilon_4 )
    \nonumber \\
    &\times& [ |M_{12\to 34}^{\uparrow\downarrow}|^2
            ({\bf u}_1-{\bf u}_2)\cdot({\bf p}_1-{\bf p}_2)   \nonumber\\
    && \quad  +  |M_{12\to 34}^{\uparrow\uparrow}|^2
            ({\bf u}_1-{\bf u}_2)\cdot {\bf q} ] \, , \label{BE4}
\end{eqnarray}
where $M^{\uparrow\downarrow}$ and $M^{\uparrow\uparrow}$ are the
amplitudes for interacting with and without spin-flip respectively.
Without spin-flip the usual factor ${\bf q}$ as in flavor diffusion appears.
With spin-flip, however, ${\bf u}_3={\bf u}_2$ and ${\bf u}_4={\bf u}_1$
and the factor $({\bf p}_1-{\bf p}_2)$ appears.
Due to Galilei invariance both terms are necessarily proportional to the
relative flow, $({\bf u}_1-{\bf u}_2)$.

The transition current can be decomposed
into interactions via the charge and the magnetic moment
by the Gordon decomposition rule
\begin{eqnarray}
    J_\mu
       &=&   \frac{g}{2m} \bar{u}_f\left[
          (p_f+p_i)_\mu +i\sigma_{\mu\nu}(p_f-p_i)^\nu
          \right] u_i \, ,
\end{eqnarray}
where only the latter can lead to spin-flip.
We notice that the spin-flip amplitude
is suppressed by a factor $p_f-p_i$ which leads to a
spin-flip amplitude suppressed by a factor $q^2$. We then find that
the spin-flip interactions do not contribute to collisions to leading
logarithmic order and the collision integral is similar to that for flavor
diffusion evaluated above. Consequently, the corresponding quark spin
diffuseness parameter becomes
\begin{eqnarray}
    D_{\sigma}^{(q)} = D_{flavor}  \, .\label{Ds}
\end{eqnarray}
Gluon spin
diffusion is slower by a factor 4/9, due to the stronger interactions,
and by another factor 4/9, due to differences between Bose and Fermi
distribution function, i.e.
$D_{\sigma}^{(g)}\simeq (4/9)^2 D_{\sigma}^{(q)}$.

Finally, let us, like for the spin diffusion, assume that color has been
polarized spatially given by a color chemical potential, $\mu_c({\bf r})$. The
basic difference to flavor and spin diffusion is that {\it quarks and gluons
can easily flip color directions} in forward scattering by color exchanges,
i.e., one does not pay the extra $q^2$ penalty factor as in the case of
spin-flip. Consequently, the color-flip interactions will dominate the
collisions since they effectively reverse the color currents.  The Boltzmann
equation thus gives us an analogous result to Eq. (\ref{BE4}) replacing spin by
color where the color-flip amplitude now dominates.  The flow velocity of the
scatterers, ${\bf u}_i$, i=1...4, depends on what color combination of the
scattering quarks, antiquarks and gluons.  However, in c.m.s.  the scatterer
has vanishing flow velocity, ${\bf u}_2=0$, on average. Likewise the final
velocities will be zero on average.  Multiplying both sides with ${\bf p}_1$
and summing the Boltzmann equation reduces to (c.f., Eq.(\ref{BE4}))
\begin{eqnarray}
    n_1&\nabla&\mu_{1,c}
    =\,  -{\bf u}_1\pi\nu_2\sum_{{\bf q},{\bf p}_1,{\bf p}_2}
   n^0_1n^0_2(1-n^0_3)(1\pm n^0_4)   \nonumber\\
    &\times&  |M_{12\to 34}|^2 ({\bf p}_1-{\bf p}_2)^2
    \delta (\varepsilon_1 +\varepsilon_2 -\varepsilon_3 -\varepsilon_4 )
    , \label{BE5}
\end{eqnarray}
where we have used the antisymmetry by interchange of ${\bf p}_1\to {\bf p}_2$.
The matrix element entering in (\ref{BE5}) is now averaged over all color
combinations.

The transverse interactions actually diverge for small momentum and energy
transfers even when integrating over
energy transfers, i.e. dynamical screening is insufficient for
obtaining a non-zero color diffusion coefficient like for the
quasiparticle decay rates in QCD and QED plasmas (see \cite{HP}).
 Concentrating therefore on the leading contribution from
transverse interactions for small $\mu=\omega/q$,
where $\Pi_T\simeq i(\pi/4)q_D^2\mu$, we find to leading order
\begin{eqnarray}
    n_1 \nabla\mu_{1,c} &=&-{\bf u}_1\nu_1 \frac{11\pi^3}{3^35\xi(3)^2}
    \alpha_s^2 T^7 \nonumber\\
   &&\times\, \int^{\sim T}_{\lambda} qdq\int^1_{-1} d\mu
    \frac{1}{q^4+(\pi/4)^2 q_D^4\mu^2}   \nonumber\\
    &=& -{\bf u}_1\nu_1 \frac{22\pi^3}{3^35\xi(3)^2}
      \alpha_s^2\frac{T^7}{q_D^2} \ln(\frac{q_D^2}{\lambda^2})
            \, , \label{q3}
\end{eqnarray}
where we have introduced an infrared cutoff, $\lambda$.
The upper limit on momentum transfers, $\sim T$, actually comes from the
distribution functions in Eq. (\ref{BE5}) but it does not enter here because
only $q\raisebox{-.5ex}{$\stackrel{<}{\sim}$} q_D$
contribute to (\ref{q3}) to leading order.
We find a color diffuseness parameter defined by
${\bf j}_c={\bf u}_1\rho_c\equiv -D_{color}\nabla\rho_c$
\begin{eqnarray}
    D_{color}^{-1} = \frac{22\pi^6}{3^65\xi(3)^2}
     \frac{1+7N_f/33}{1+N_f/6}
     \, \alpha_s\ln(q_D^2/\lambda^2) T \, .
\end{eqnarray}
With $\lambda\sim g^2T$ the analogous result to Eq. (\ref{gg}) is obtained
\begin{eqnarray}
    D_{color}^{-1} \simeq 4.9 \, \alpha_s\ln(1/\alpha_s) T \, ,\label{Dc}
\end{eqnarray}
(where we have ignored the minor dependence on $N_f$).
Comparing Eqs. (\ref{Ds}) and (\ref{Dc}) we see that
$D_{color} \sim \alpha_s D_\sigma=\alpha_s D_{flavor}$.
The color-flip mechanism amplifies the forward collisions so the color cannot
diffuse through the quark-gluon plasma as easily as spin or flavor.

The factor $\ln(1/\alpha_s)$ in $D_{color}$ has a completely different origin
as the one in $D_{flavor}$ or $D_{spin}$. In $D_{color}$ the logarithm arises
from an integral $dq/q$ over momentum transfers from $q\sim\lambda\sim g^2T$ to
$q\sim q_D\sim gT$ as in the case of quark and gluon quasiparticle decay rates
\cite{HP}.  In $D_{flavor}$ or $D_{spin}$ and the transport coefficient
discussed in \cite{BP1,BP2} a similar integral occurs, but with momentum
transfers from $q\sim q_D\sim gT$ to $q\sim T$. Thus the infrared cutoff does
not enter these transport coefficients.  In both cases the result is
proportional to the logarithm of the ratio of the upper and lower limits on the
momentum transfer, namely $\ln(1/g)$. The difference in the important range of
momentum transfers in the two cases is due to the absence in the calculation of
$D_{color}$ of the extra factor $\sim q^2/T^2$.  Therefore small momentum
transfer processes have greater weight in the calculation of $D_{color}$ and
quasiparticle relaxation rates than they do for $D_{flavor}$, $D_{spin}$ and
standard transport relaxation rates. In the latter cases the factor
$q^2/T^2\sim q_D^2/T^2\sim \alpha_s$ also reduces the rates by a factor
$\alpha_s$.

   Other related transport coefficients are the electrical conductivity,
$\sigma_{el}$, in QED and the corresponding color conductivity,
$\sigma_{color}$, in QCD.
Applying a color-electric field, ${\bf E}_c$, to the quark-gluon plasma
generates a color current, ${\bf j}_c$. The color conductivity
$\sigma_{color}=-j_c/E_c$ can thus be found by solving the Boltzmann equation
analogous to the color diffusion process. We find
\begin{eqnarray}
    \sigma_{color} = \frac{2}{3} g^2 D_{color}
   \sum_{i,{\bf p}}  \nu_i\left(\frac{\partial n_i}{\partial\varepsilon_i}
  \right)     \, .\label{Sc}
\end{eqnarray}
Here $D_{color}$ plays the role of the color relaxation time.
Eq. (\ref{Sc}) is the standard result for a plasma except for the factor 2/3
which arises because only two-thirds of the colors contribute to the currents
for a given color field. Inserting $D_{color}$ from (\ref{Dc}) we obtain
\begin{eqnarray}
    \sigma_{color}
   &\simeq& \frac{8\pi}{3}N_f \alpha_s D_{color} T^2
   \, \simeq  1.7 N_f T/\ln(1/\alpha_s) \, ,\label{sc}
\end{eqnarray}
from quark currents alone. Gluon currents are slower due to stronger
interactions and will reduce the conductivity slightly.
This result differs from \cite{GS} by a numerical factor only.
For comparison the electrical conductivity below
$T\raisebox{-.5ex}{$\stackrel{<}{\sim}$} M_W$ is
$\sigma_{el}\sim T/\alpha\ln(1/\alpha)$ \cite{BP2}. The different
dependence on coupling constant arises because the exchanged photon
does not carry charge whereas the exchanged gluon can carry color.
The characteristic relaxation times for conduction are very
different in QCD, where
$\tau_{color}\sim D_{color}\sim (\alpha_s\ln(1/\alpha_s)T)^{-1}$,
as compared to QED, where
$\tau_{el}\simeq (\alpha^2\ln(1/\alpha)T)^{-1}$.
Consequently, QGP are much poorer color conductors than QED plasmas when
$T\ll M_W$ for the same coupling constant.

These surprising results for QCD are qualitatively in agreement with those
found by Selikhov \& Gyulassy \cite{GS} who have considered the diffusion
of color in color space. They use the fluctuation-dissipation
theorem to estimate the deviations from equilibrium and find the same
color non-flip and color-flip terms, which they denote the momentum and
color diffusion terms respectively,
and they also find that the latter dominates being infrared divergent.
Inserting the same infrared cut-off they find a color diffusion
coefficient in color space equal to Eq. (\ref{gg})
\begin{eqnarray}
    d_c = \frac{1}{\tau^{(g)}_1}
        =  3\alpha_s\ln(1/\alpha_s) T \, . \label{dc}
\end{eqnarray}
Note that this quantity is proportional to the inverse of $D_{color}$
as given in Eq. (\ref{Dc}).

The color-flip mechanism is not restricted to QCD but has analogues in
other non-abelian gauge
theories. In the very early universe when $T\gg M_W\simeq 80$ GeV, the
$W^\pm$ bosons can be neglected and faces the same
electroweak screening problems as QCD and QED. Since
now the exchanged $W^\pm$ bosons carry charge (unlike the photon, but
like the colored gluon), they can easily change the charge of, for example,
an electron to a neutrino in forward scatterings. Thus
the collision term will lack the usual factor $q^2$  as for the
quasiparticle damping rate and the color diffusion.
Since $SU(2)\times U(1)$ gauge fields should have the same
infrared problems as SU(3) at the scale of the magnetic mass, $\sim e^2 T$,
we insert this infrared cutoff. Thus we find a diffusion parameter
for charged electroweak particles in the very early universe of order
\begin{eqnarray}
   D_{el} \sim (\alpha \ln(1/\alpha) T)^{-1}  ,
\end{eqnarray}
which  is a factor $\alpha$ smaller than when $T\ll M_W$.
The electrical conductivity will be smaller
by the same factor as well, $\sigma_{el}\sim T/\ln(1/\alpha)$.

In summary, the flavor, spin and color diffusion coefficients have been
calculated in QCD plasmas to leading order in the interaction strength.  Color
diffusion and the quark and gluon quasiparticle decay rates are not
sufficiently screened and do depend on an infrared cut-off of order the
magnetic mass, $m_{mag}\sim g^2T$; typically $D_{color}^{-1}\sim
\alpha_s\ln(q_D/m_{mag})T \sim \alpha_s\ln(1/\alpha_s)T$.  Flavor and spin
diffusion processes are sufficiently screened by Debye screening for the
longitudinal or electric part of the interactions and by Landau damping for the
transverse or magnetic part of the interactions; typically
$D_{flavor}^{-1}=D_{spin}^{-1}\sim\alpha_s^2\ln(1/\alpha_s)T$.  As a
consequence, color diffusion is slow and the QGP is a poor color conductor. In
the very early universe when $T\gg M_W$ exchanges of $W^\pm$ provides charge
exchange - a mechanism analogous to color exchange in QCD - and QED plasmas
will also be poor electrical conductors.

This work was supported in part by DOE grant No. DE-AC03-76SF00098 and the
Danish Natural Science Research Council.  Discussions with Gordon Baym, Chris
Pethick and Alexei Selikhov are gratefully acknowledged.


\begin{references}
\bibitem{GS} A. Selikhov and M. Gyulassy, {\em Phys. Lett.}
             {\bf B316}, 316 (1993); and CU-TP-610/93.
\bibitem{BP1} G. Baym, H. Monien and C. J. Pethick, Proc.
    {\it XVI Int.  Workshop
    on Gross Properties of Nuclei and Nuclear Excitations}, Hirschegg,
    (ed. H. Feldmeier, GSI and Institut f{\" u}r Kernphysik, Darmstadt, 1988),
    p. 128; C. J. Pethick,
    G. Baym and H. Monien, {\it Nucl.  Phys.}  {\bf A498}, 313c (1989);
    G. Baym, H. Monien, C. J. Pethick, and D. G. Ravenhall, {\it Phys. Rev.
    Lett.} {\bf 64}, 1867 (1990).
\bibitem{BP2} G. Baym, H. Monien, C. J. Pethick, and D. G. Ravenhall,
    {\it Nucl. Phys.} {\bf A525}, 415c (1991);  G. Baym, H. Heiselberg,
    C. J. Pethick, and J. Popp, {\it Nucl. Phys.} {\bf A544}, 569c (1992).
\bibitem{GT}  M. Gyulassy and M. Thoma, Nucl. Phys. {\bf B 351}, 491 (1991);
   E. Braaten and M. Thoma, Phys. Rev.  {\bf D44}, 1298, R2625 (1991)
\bibitem{damp} C. P. Burgess and A. L. Marini, {\it Phys.  Rev.} {\bf
      D45}, R17 (1992); A. Rebhan, {\it Phys.  Rev.} {\bf D48}, 482 (1992);
      R. D. Pisarski, {\it Phys. Rev.} {\bf D47}, 5589 (1993).
\bibitem{HP} H. Heiselberg and C. J. Pethick, {\it Phys. Rev.} {\bf D47},
             R769 (1993);
\bibitem{deg} H. Heiselberg, G. Baym, and C. J. Pethick, {\it Nucl.
               Phys.  B} (Proc.  Suppl.)  {\bf 24B} 144, (1991);
               H. Heiselberg and C. J. Pethick,
               {\it Phys. Rev.} {\bf D48}, 2916 (1993).
\bibitem{LH} H. Heiselberg and C.J. Pethick, NBI-93-19,
           proceedings of {\em ``Plasma Physics"},
           Les Houches, Feb. 2-11, 1993.
\bibitem{We}  H. A. Weldon, {\it Phys. Rev.} {\bf D26}, 1394 (1982).
\bibitem{Linde} A. D. Linde, {\it Phys. Lett.} {\bf B96}, 289 (1980).
\bibitem{BPbook} G. Baym and C. J. Pethick, {\it Landau Fermi-liquid theory:
            concepts and applications} (Wiley, New York, 1991).
\end{references}
\end{document}